\documentclass[
aps,%
10pt,%
final,%
notitlepage,%
oneside,%
onecolumn,%
nobibnotes,%
nofootinbib,
superscriptaddress,%
noshowpacs%
,centertags%
]%
{revtex4}
\usepackage[cp1251]{inputenc}
\usepackage[T2A]{fontenc}
\usepackage[english,russian]{babel}
\usepackage{amsfonts}
\usepackage{amsbsy}
\usepackage{mathrsfs}
\usepackage{feynmp}
\usepackage{graphicx}
\let\sectiontmp\section\let\subsectiontmp\subsection \let\appendixtmp\appendix
\def\section{\setcounter{equation}{0}\sectiontmp}
\def\subsection{\subsectiontmp}
\def\theequation{\arabic{section}.\arabic{equation}}
\def\appendix{\def\theequation{\Alph{section}.\arabic{equation}}\appendixtmp}

\def\contourxy{\unitlength=0.8cm
\begin{picture}(17,1)\thicklines
\contour
\put(14.5,-.5){\makebox(0,0){$\infty$}}
\put(11,-.5){\makebox(0,0){$t_x^+$}}
\put(8,1.8){\makebox(0,0){$t_y^-$}}
\put(11,0){\circle*{.2}}
\put(8,1){\circle*{.2}}
\end{picture}}
\def\contour{\thicklines
\put(1,-.5){\makebox(0,0){$t_{0}$}}
\put(16.5,.5){\makebox(0,0){$t$}}
\put(0,.5){\vector(1,0){16}}
\put(1,1.){\line(1,0){13}}\put(14,.5){\oval(1,1)[br]}
\put(14,0){\vector(-1,0){13}}\put(14,.5){\oval(1,1)[tr]}}

\linethickness{1.0pt}
\def\geight{\begin{picture}(20,0)
\put(0,2){
     \thicklines
     \put(10,0){\circle*{2.00}}
     \put(10,7.5){\circle{15.05}}
     \put(10,-7.5){\circle{15.05}}}
     \end{picture}}
\def\sloop{\begin{picture}(20,0)
\put(0,2){
     \thicklines
     \put(10,0){\circle*{4.00}}
     \put(10,7.5){\circle{15.00}}}
     \end{picture}}
\def\GGJ{\begin{picture}(20,15)
\put(0,2){
     \thicklines
     \put(10,0){\circle*{2.00}}
     \put(10,7.5){\circle{15.00}}
     \put(10,7.5){\makebox(0,0){$\ii J$}}
     \put(10,0){\thinlines\line(0,-1){10}}}
     \end{picture}}
\def\GPb{\begin{picture}(10,0)
\put(0,2){\put(5,3){\thinlines\line(0,-1){6}}
          \put(5,3){\makebox(0,0){$\times$}}}
     \end{picture}}
\def\GPbfull{\begin{picture}(10,0)
\put(0,2){\put(5,3){\line(0,-
1){6}}\put(5,3){\makebox(0,0){$\bigotimes$}}}
     \end{picture}}
\def\GG{\begin{picture}(24,0)
\put(22,2){\thinlines\vector(-
1,0){12}}\put(2,2){\thinlines\line(1,0){10}}
     \end{picture}}
\def\GGfull{\begin{picture}(24,0)
\put(22,2){\thicklines\vector(-
1,0){12}}\put(2,2){\thicklines\line(1,0){10}}
     \end{picture}}

\def\Dysonself{\begin{picture}(60,0)
\put(0,2){\thicklines
     \put(0,0){\thinlines\line(1,0){10}}
        \put(20,0){\thinlines\vector(-1,0){12}}
     \put(60,0){\vector(-1,0){12}}\put(40,0){\line(1,0){10}}
     \put(30,0){\oval(20,15)}
     \put(30,0){\makebox(0,0){$-\ii\Se$}}}
     \end{picture}}
\def\gloopphitwo{\begin{picture}(20,0)
\put(0,2){
     \thicklines
     \put(10,0){\line(-1,-1){6}}
     \put(10,0){\line(1,-1){6}}
     \put(10,0){\circle*{2.00}}
     \put(4,-6){\makebox(0,0){$\bigoplus$}}
     \put(16,-6){\makebox(0,0){$\bigoplus$}}
     \put(10,7.5){\circle{15.00}}}
     \end{picture}}
\def\gloopphione{\begin{picture}(20,0)
\put(0,2){
     \thicklines
     \put(10,0){\line(0,-1){6}}
     \put(10,0){\circle*{4.00}}
     \put(10,-6){\makebox(0,0){$\bigotimes$}}
     \put(10,7.5){\circle{15.00}}}
     \end{picture}}
\def\gphifour{\begin{picture}(22,0)
\put(4,2){
     \thicklines
     \put(2,5){\line(1,-1){10}}
     \put(12,5){\line(-1,-1){10}}
     \put(7,0){\circle*{2.00}}
     \put(2,-5){\makebox(0,0){$\bigoplus$}}
     \put(12,-5){\makebox(0,0){$\bigoplus$}}
     \put(2,5){\makebox(0,0){$\bigoplus$}}
     \put(12,5){\makebox(0,0){$\bigoplus$}}}
     \end{picture}}
\def\sphitwo{\begin{picture}(22,0)
\put(4,2){
     \thicklines
     \put(7,0){\line(1,1){6}}
     \put(4.5,0){\line(1,0){5}}
     \put(7,0){\line(-1,1){6}}
     \put(7,0){\circle*{4.00}}
     \put(2,5){\makebox(0,0){$\bigoplus$}}
     \put(12,5){\makebox(0,0){$\bigoplus$}}}
     \end{picture}}
\def\gphithree{\begin{picture}(22,0)
\put(4,2){
     \thicklines
     \put(2,0){\line(1,0){12}}
     \put(8,0){\line(0,1){6}}
     \put(8,0){\circle*{4.00}}
     \put(14,0){\makebox(0,0){$\bigotimes$}}
     \put(8,6){\makebox(0,0){$\bigotimes$}}
     \put(2,0){\makebox(0,0){$\bigotimes$}}}
     \end{picture}}
\def\gsand4{\begin{picture}(38,0)
\put(2,2){
     \thicklines
     \put(2,0){\circle*{2.00}}
     \put(32,0){\circle*{2.00}}
     \bezier{216}(2,0)(17,-10)(32,0)
     \bezier{216}(2,0)(17,+10)(32,0)
     \bezier{284}(2,0)(17,-25)(32,0)
     \bezier{284}(2,0)(17,+25)(32,0)}
     \end{picture}}
\def\ssand3{\begin{picture}(38,0)
\put(2,2){
     \thicklines
     \put(2,0){\line(1,0){30}}
     \put(2,0){\circle*{4.00}}
     \put(32,0){\circle*{4.00}}
     \bezier{284}(2,0)(17,-25)(32,0)
     \bezier{284}(2,0)(17,+25)(32,0)}
     \end{picture}}
\def\sphisand2phi{\begin{picture}(42,0)
\put(2,2){
     \thicklines
     \put(2,0){
     \put(2,0){\line(1,0){8}}
     \put(24,0){\line(1,0){8}}
     \put(10,0){\makebox(0,0){$\bigotimes$}}
     \put(24,0){\makebox(0,0){$\bigotimes$}}
     \put(2,0){\circle*{4.00}}
     \put(32,0){\circle*{4.00}}
     \bezier{284}(2,0)(16,-25)(32,0)
     \bezier{284}(2,0)(16,+25)(32,0)}}
     \end{picture}}
\def\gphisand3phi{\begin{picture}(53,0)
\put(2,2){
     \thicklines
     \put(3,0){\line(1,0){44}}
     \put(3,0){\makebox(0,0){$\bigotimes$}}
     \put(47,0){\makebox(0,0){$\bigotimes$}}
     \put(8,0){
     \put(2,0){\circle*{2.00}}
     \put(32,0){\circle*{2.00}}
     \bezier{284}(2,0)(16,-25)(32,0)
     \bezier{284}(2,0)(16,+25)(32,0)}}
     \end{picture}}
\def\gloopphih{\begin{picture}(43,0)
\put(2,2){
     \thicklines
     \put(2,0){\line(1,0){37}}
     \put(39,0){\makebox(0,0){$\bigotimes$}}
     \put(2,0){\circle*{4.00}}
     \put(32,0){\circle*{2.00}}
     \bezier{284}(2,0)(16,-25)(32,0)
     \bezier{284}(2,0)(16,+25)(32,0)}
     \end{picture}}
\def\DSa{\begin{picture}(0,0)\thicklines
\put(0,0){\oval(1.5,1)}
\put(0,0){\makebox(0,0){$-\ii\Sa$}}\end{picture}}
\def\GlnG0Sa{
\begin{picture}(4.3,1.5)
\put(0.5,.1){
\put(.5,0){\DSa}\put(3,0){\DSa}\put(1.75,1){\DSa}
\put(1.75,-1){\makebox(0,0){\dots\dots}}
\put(1,.5){\oval(1,1)[lt]}\put(2.5,.5){\oval(1,1)[rt]}
\put(1,-.5){\oval(1,1)[lb]}\put(2.5,-.5){\oval(1,1)[rb]}}
\end{picture}}
\def\GGaSa{
\begin{picture}(2.5,2.)
\thicklines\put(0.5,.1){
\put(.5,0){\DSa}\put(2,-.5){\line(0,1){1}}
\put(1.,1){\line(1,0){0.5}}\put(1.,-1){\line(1,0){0.5}}
\put(1,.5){\oval(1,1)[lt]}\put(1.5,.5){\oval(1,1)[rt]}
\put(1,-.5){\oval(1,1)[lb]}\put(1.5,-.5){\oval(1,1)[rb]}}
\end{picture}}
\def\vhight#1{\vphantom{\left(\begin{picture}(0,#1)\end{picture}\right)}
}
\def\Dclosed#1#2{
\begin{picture}(2,.8)\put(0,.1){#2
\put(1,0){\circle{1.414}}
\put(1,-.707){\line(0,1){1.414}}\put(.293,0){\line(1,0){1.414}}
\put(0.5,-0.5){\line(0,1){1}}\put(1.5,-0.5){\line(0,1){1}}
\put(0.5,-0.5){\line(1,0){1}}\put(0.5,0.5){\line(1,0){1}}}
\put(1.5,-.8){$#1$}
\end{picture}}
\def\Dclosedone#1#2{
\begin{picture}(2,.8)\put(0,.1){#2
\put(1,0){\circle{1.414}}
\put(1,-.707){\line(0,1){1.414}}\put(.293,0){\line(1,0){1.414}}
\put(0.5,-0.5){\line(0,1){1}}\put(1.5,-0.5){\line(0,1){1}}
\put(0.5,-0.5){\line(1,0){1}}\put(0.5,0.5){\line(1,0){1}}}
\put(1.5,-.8){$#1$}
\put(.293,0.1){\circle*{.2}}
\end{picture}}
\newcommand{\di}{{\mathrm d}}
\newcommand{\Tr}{{\mathrm{Tr}}}
\newcommand{\ii}{{\mathrm i}}

\renewcommand{\and}{\quad{\mathrm{and}}\quad}

\renewcommand{\oint}{\int_{\cal C}}
\renewcommand{\Im}{{\mathrm{Im}}}
\def\Tc{{\cal T}_{\cal C}}

\def\scr#1{\mbox{\scriptsize #1}}

\def\vec#1{\mbox{\boldmath $#1$}}
\newcommand{\dpi}[1]{\frac{\di^4 #1}{(2\pi)^4}}

\newlength{\charwidth}
\def\medhat#1{\settowidth{\charwidth}{$#1\,$}{\makebox[\charwidth]{$\,
 {\widehat{\makebox[2mm]{$#1\,$}}}$}}\vphantom{#1}}
\newcommand{\lap}%
{\raisebox{-0.5ex}{$\stackrel{\scriptstyle <}{\scriptstyle \sim}$}}
\newcommand{\gap}%
{\raisebox{-0.5ex}{$\stackrel{\scriptstyle >}{\scriptstyle \sim}$}}

\def\Gr{\Delta}\def\Se{\Pi}
\def\j{\medhat{J}}
\def\ja{\medhat{J}}\def\jad{\medhat{J}^\dagger}

\def\Pa{\medhat{\phi}}\def\Pad{\Pa^\dagger}
\def\Pta{\medhat{\varphi}}
\def\Pt{\medhat{\varphi}}\def\Ptd{\Pt^\dagger}

\def\Ph{\medhat\phi}\def\Phd{\medhat{\phi}^{\dagger}}

\def\Pba{{\phi}}\def\jba{{J}}
\def\dc{\delta_{\cal C}}
\def\Ga{\Gr}
\def\Sa{\Se}

\def\A{A}
\def\Gm{\Gamma}
\def\suma{\displaystyle \left(\frac{1}{2}\right)_{\scr{neut.}}}
\def\Lg{{\cal L}}
\def\Lgh{\makebox[3.5mm]{${\widehat{\makebox[2mm]{$\Lg$}}}$}\vphantom{L}}
\def\Lint{\Lgh^{\mbox{\scriptsize int}}}

\def\Hh{\medhat{H}}

\def\So{S}

\begin{document}
\selectlanguage{english}  
\title{Non-Equilibrium 2PI Potential
and Its Possible Application to Evaluation of
Bulk Viscosity\footnote{Dedicated to
S.T. Belyaev on the occasion of his 90th birthday.}}

\author{Yu.~B.~Ivanov}
\email[]{Y.Ivanov@gsi.de}
\affiliation{NRC ``Kurchatov Institute'', Kurchatov sq.$\!$ 1, Moscow
123182, Russia}
%
\author{D.~N. Voskresensky}
\email[]{D.Voskresensky@gsi.de}
\affiliation{National Research Nuclear University "MEPhI",
Kashirskoe sh. 31, Moscow 115409, Russia}
%

\begin{abstract}
Within non-equilibrium Green's function technique on the
real-time contour and the two-particle-irreducible (2PI) 
$\Phi$-functional method, a non-equilibrium  potential is
introduced. It naturally generalizes  the conventional
thermodynamic potential with which it coincides in thermal
equilibrium. Variations of the non-equilibrium potential over
respective parameters result in the same quantities as those of
the thermodynamic potential but in arbitrary non-equilibrium. In
particular, for slightly non-equilibrium  inhomogeneous
configurations a variation of the non-equilibrium potential over
volume is associated with the trace of the non-equilibrium stress
tensor. The latter is related to the bulk viscosity.  This
provides a novel way for evaluation of the bulk viscosity.
\end{abstract}
\maketitle

\section{Introduction}

Non-equilibrium Green's function technique, developed by
Schwinger, Kadanoff, Baym and Keldysh
\cite{Schw,KadB,Kad62,Keld64,Lif81}, is the appropriate concept to
study the space--time evolution of many-particle quantum systems.
This formalism finds now applications in various fields, such as
quantum chromo-dynamics \cite{Land}, nuclear physics
\cite{Dan84,Toh,Bot90,MSTV,Knoll95,IKV,IKV2,IKHV,KIV01,IKV3,Hees,IV09,V2011},
astrophysics \cite{MSTV,VS87,Keil}, cosmology
\cite{CalHu,KKL2009}, spin systems \cite{Manson}, lasers
\cite{Korenman},  plasma physics \cite{Bez,Kraft}, physics of
liquid $^{3}$He \cite{SerRai}, critical phenomena, quenched random
systems and disordered systems \cite{Chou}, normal metals and
super-conductors \cite{VS87,Rammer,Fauser}, semiconductors
\cite{LipS}, tunneling and secondary emission \cite{Noziers},
ultra-cold gases \cite{KG2011}, etc. This list is certainly not
complete.

With the aim to come to a tractable dynamical scheme one
compulsory performs partial re-summations.  In contrast to
perturbation theory, however, doing re-summations one frequently
encounters a complication, that the resulting equations of motion
may no longer comply with the conservation laws, e.g., of charge,
energy and momentum. This problem was considered in
\cite{Baym}
 in terms of equilibrium Green's function formalism. Any
approximation, in order to be conserving, must be so-called
$\Phi$-derivable, see  also \cite{Luttinger,Abrikos}. Any
$\Phi$-derivable approximation provides thermodynamically
consistent results.  A common generating functional depending on
auxiliary two-particle irreducible (2PI) $\Phi$ functional was
then constructed on equilibrium \cite{Cornwall} and
non-equilibrium \cite{IKV} Schwinger-Keldysh contours, being
determined in terms of full, i.e. re-summed, classical fields and
Green's functions coupled by free vertices. The presented scheme
of constructing self-consistent approximations provides a suitable
basis for the derivation of generalized kinetic description beyond
the limitations of the quasi-particle approximation
\cite{IKV2,KIV01,IKHV,IKV3,Hees,IV09}. Such generalized transport
schemes respect the quantum nature of the particles  and, that is
of particular importance, take into account  their finite
mass-widths. Relativistic effects first introduced in quantum
kinetics by Belyaev and Budker \cite{BB} are easily incorporated
in the mentioned general scheme.

 Close to equilibrium, at time intervals much larger than the
time scale of changes of kinetic quantities the generalized
kinetic description can be replaced by a more economical
hydrodynamical
 description. The fluid-dynamical approach is fairly efficient for
description of heavy-ion collisions in a broad incident energy
range from SIS to LHC energies.
Interest in the transport coefficients  within the hydrodynamics
has been raised by large values of  elliptic flow at RHIC
energies \cite{RHIC-v2}, though their study started long before
\cite{Gal79,DanPL84}. The observed   elliptic flow indicated that
the created quark-gluon plasma behaves like a fluid with a  low
(but non-zero) value of the shear viscosity \cite{Romatschke}.
The bulk viscosity can also essentially affect the dynamics of
heavy-ion collisions, supernovas in astrophysics and acceleration
of the Universe in cosmology \cite{Gagnon}. As it was shown in
\cite{Torrieri:2008ip}, large value of the bulk viscosity can
result in instability at the freeze-out stage. A contribution to
the pressure related to the bulk viscosity affects the equation of
state, making it softer for expanding systems and stiffer for
contracting ones \cite{V2011}.  The stability of   neutron
stars to the growth of  $r$-modes is provided only by large values
of the bulk viscosity of the matter \cite{Anderson}. 
Transport coefficients also  govern  the dynamics of the
first-order phase transitions \cite{SV09}.

Evaluation of transport coefficients of strongly
interacting matter is usually done within the
quasiparticle approximation ~\cite{SHMCbulk}, although the
width effects can be large at least for some species.
Studies of transport coefficients for resonances have  just recently  started
\cite{V2011,BS06}. Moreover, all above derivations were
performed within simplified approximation, i.e. the relaxation time approximation.
A rigorous way to evaluate transport coefficients is based on
Kubo formulas \cite{Kubo}. Calculations of such kind are  highly complicated
\cite{Jeon,Filinov:2012dp}.
In \cite{Aarts:2003bk} the problem
was discussed in the framework of the 2PI method. The attention
was primarily focused on the shear viscosity. In the present paper
we mainly address the bulk viscosity as one of the possible
applications of the introduced non-equilibrium 2PI potential. 
The case of the bulk viscosity is extra complicated. In addition to the
conventional kinetic contribution related to single-particle dynamics,
the bulk viscosity contains an important contribution coming from fluctuations of
soft collective modes \cite{ML37,LL06}. The latter contribution appreciably
increases  in the vicinity of the phase transition 
\cite{Kharzeev}. Therefore, a reliable scheme for calculating the bulk viscosity
should properly incorporate both these contributions.

The aim of the present paper is introduction of a non-equilibrium potential
that generalizes the conventional thermodynamic potential beyond the scope of
equilibrium. We describe one of the possible applications of this new potential,
i.e. its application to evaluation of the bulk viscosity proceeding from
$\Phi$-derivable approximations within non-equilibrium Green's function technique.
In sect.
\ref{Prerequisites} we  briefly formulate concepts developed in
\cite{IKV}. In sect. \ref{Potential} we derive a non-equilibrium potential,
based on the generating functional,  and
show that the non-equilibrium potential transforms into the conventional
thermodynamic potential in thermal equilibrium.
 Then in sect. \ref{bulk}  we discuss  relations between the non-equilibrium
 potential, the pressure,
 the stress tensor and the bulk viscosity.
 Some helpful relations are deferred to the Appendix \ref{discontinouity}.


\section{Prerequisites}\label{Prerequisites}
\label{sect-Prel} In this section we briefly formulate concepts
developed in \cite{IKV} and introduce necessary notations.
\subsection{Noether Energy-Momentum Tensor and Current}\label{Noether-Tensor}
 For notational convenience we consider a system
of {\em
  relativistic} scalar bosons, specified by free Klein-Gordon Lagrangians
\begin{eqnarray}
\Lgh^0 =
\left\{
\begin{array}{ll}
\frac{1}{2}\displaystyle\vphantom{\frac{1}{m}}
 \left(\partial_\mu\Pa \cdot \partial^\mu\Pa
-  m^2 (\Pa)^2\right)
\quad&{\mbox{for neutral bosons}}\\[2mm]
\displaystyle
\left(
\partial_\mu\Pad\cdot\partial^\mu\Pa
- m^2 \Pad\Pa\right)
\quad&{\mbox{for charged bosons}}
\end{array}\right.
\label{L0cb}
\end{eqnarray}
%
where $\Pa(x)$ and $\Pad(x)$ are bosonic field operators. All
considerations are straight forwardly  adapted to multi-component
systems of different flavors, to fields with intrinsic degrees
of freedom, to non-scalar fields, such as vector boson or Dirac
fermion fields or also to non-relativistic many-body pictures. The
interaction Lagrangians $\Lint\{\Pa\}$ (for neutral bosons) and
the charge symmetric $\Lint\{\Pa,\Pad\}$ for charged bosons are
assumed to be local, i.e.  without derivative coupling.
Generalization to the derivative coupling interaction can be
performed following the line of \cite{IKV3}.

The variational principle of stationary action leads to the Euler--Lagrange
equations of motion for the field operators
%
\begin{eqnarray}
\label{eqmotion}
\So_x \Pa(x)&=& - \ja(x)=-\frac{\partial
\Lint}{\partial\Pad},\quad\quad\mbox{where }\quad
\So_x=
-\partial_\mu\partial^\mu -m^2 ,
\end{eqnarray}
%
and similarly for the corresponding adjoint equation.  Thereby the
$\ja(x)$ operator is a local source current of the field $\Pa$,
while $\So_x$ is the differential operator of the free evolution
with the free propagator $\Ga^0(y,x)$, as resolvent.

The canonical form \cite{Itz80} of the energy-momentum tensor
operator reads
%
\begin{equation}
\label{E-M-standard-tensor}
\medhat{\Theta}^{\mu\nu}_{\scr{(canon.)}} =
\kappa \left( \frac{\partial
\Lgh}{\partial\left(\partial_\mu\Pa\right)}
\partial^\nu\Pa
+ \frac{\partial
\Lgh}{\partial\left(\partial_\mu\Pad\right)}\left(\partial^\nu\Pad\right)
\right) -g^{\mu\nu}\Lgh,
\end{equation}
%
which is conserved, i.e.
%
$\partial_\mu {\widehat \Theta}^{\mu\nu}_{\scr{(canon.)}} = 0$.
%
Here  and below $\kappa =1/2$ for neutral bosons and $1$ for
charged bosons, $g^{\mu\nu}=\mbox{diag}(1,-1,-1,-1)$. The
canonical energy-momentum tensor  is non-symmetric. Alternatively,
using equations of motion
(\ref{eqmotion}) one can derive symmetric expression for the
conserved energy momentum tensor.
Introducing convenient notation $\widehat{p}^{\mu}_x = \ii
\partial^\mu_x$ we rewrite the symmetric energy--momentum tensor
as
%
\begin{eqnarray}
\label{E-M-pi-tensor} \nonumber \widehat{\Theta}^{\mu\nu}(x) &=&
\frac{1}{4}\kappa\left[ \left(\left(\widehat{p}^{\mu}_x\right)^* +
\widehat{p}^{\mu}_y\right)\left(\left(\widehat{p}^{\nu}_x\right)^*
+ \widehat{p}^{\nu}_y\right) \left[\Pad(x),\Pa(y)\right]
\right]_{x=y}\\ &+& g^{\mu\nu} \left({\widehat{\cal
E}}^{\scr{int}}(x)- {\widehat{\cal E}}^{\scr{pot}}(x) \right) .
\end{eqnarray}
%

Here $[...,...]$ denotes commutator and   we  introduced
operators of the interaction-energy density ${\widehat{\cal
E}}^{\scr{int}}$ and the potential-energy density ${\widehat{\cal
E}}^{\scr{pot}}$:
%
\begin{equation}
\label{eps} {\widehat{\cal E}}^{\scr{int}}(x) = - \Lint(x),\quad
\end{equation}
%
%
\begin{eqnarray}
\label{eps-pot} {\widehat{\cal E}}^{\scr{pot}}(x)
=-\frac{1}{2}\kappa \left( \jad(x) \Pa(x) +  \ja(x) \Pad(x)
\right).
\end{eqnarray}
%
It is easy to show that
$\partial_\mu \left(\medhat{\Theta}^{\mu\nu}-
\medhat{\Theta}^{\mu\nu}_{\scr{(canon.)}} \right) =0$. This implies that
$\medhat{\Theta}^{\mu\nu}$ and
$\medhat{\Theta}^{\mu\nu}_{\scr{(canon.)}}$ are equivalent.

For specific interaction Lagrangians, e.g., with a certain number
$\gamma$ of operators attached to each vertex, eq. (\ref{eps-pot})
together with definition (\ref{eqmotion}) for the current $\ja$
allow  to obtain a simple relation between ${\widehat{\cal
E}}^{\scr{pot}}$ and ${\widehat{\cal E}}^{\scr{int}}$,
%
\begin{eqnarray}\label{gamma}
{\widehat{\cal E}}^{\scr{int}}(x) =\frac {2}{\gamma}{\widehat{\cal
E}}^{\scr{pot}}(x).
\end{eqnarray}
%
 For  $\phi^4$-theory
$\gamma=4$.


 If the Lagrangian is invariant under some global transformation
of complex fields (with the charge $e$), e.g.,
%
\begin{equation}
\label{c-global-tr.} \Pa(x)\Rightarrow e^{-\ii
e\Lambda}\Pa(x);\quad\quad \Pad(x)\Rightarrow e^{\ii
e\Lambda}\Pad(x),
\end{equation}
%
 there exists Noether
current defined as \cite{Itz80},
%
\begin{equation}
\label{c-pi-current} \medhat{j}^{\mu}_{\scr{(Noether)}}(x) = e
\frac{1}{2}\left[ \left(\left(\widehat{p}^{\mu}_x\right)^* +
\widehat{p}^{\mu}_y\right) \left[\Pad(x),\Pa(y)\right]_\mp
\right]_{x=y}.
\end{equation}
%
which is conserved, i.e.
%
$\partial_{\mu} \medhat{j}^{\mu}_{\scr{(Noether)}} = 0.$
%
This current naturally vanishes for neutral fields.

\subsection{Real-Time Contour}

In the non-equilibrium case, one assumes that the system has been
prepared at some initial time $t_0$ in terms of its density
operator $\medhat{\rho}_0=\sum_a P_a\left|a\right>\left<a\right|$
at that time, where  $\left|a\right>$ form a complete set of
eigenstates of $\medhat{\rho}_0$.

\parbox[t]{14.5cm}{
\begin{center}\vspace*{1cm}
\contourxy\\[1cm]
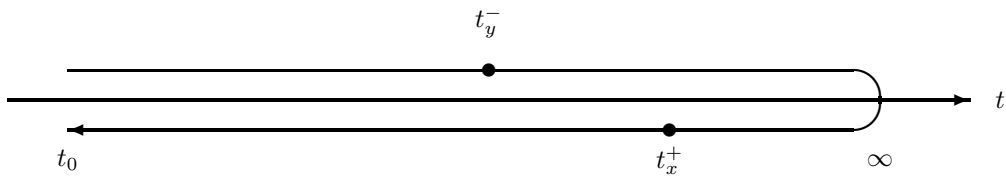
Figure: Closed real-time contour with two external points $x,y$ on the contour.
\end{center}}

The non-equilibrium theory can  be formulated on a  {\em closed
real-time contour} (see figure), where the time  runs from $t_{0}$
to $\infty$ along {\em time-ordered} branch and back to $t_{0}$
along {\em anti-time-ordered} branch. Contour-ordered multi-point
functions are defined as expectation values of contour-ordered
products of operators,
%
\begin{equation}\label{cont.exp.}
\left<\Tc \medhat{A}(x_1)\medhat{B}(x_2)\dots\right>
=\left<\Tc \medhat{A}_{\rm I}(x_1)\medhat{B}_{\rm I}(x_2)\dots
\exp\left\{\ii\oint\Lint_{\rm I}\di x\right\}\right>,
\end{equation}
%
where $\Tc$  orders the operators according to a time parameter
running along the  contour ${\cal
  C}$. The l.h.s.  is written in the Heisenberg
representation,  the r.h.s.,  in the  interaction picture
(subscript "I").

In certain calculations, e.g., in those that apply Fourier and
Wigner transformations, it is necessary to decompose the full
contour into its two branches---the {\em time-ordered} and {\em
anti-time-ordered} branches. One then has to distinguish between
the physical space-time coordinates $x,\dots$ and the
corresponding contour coordinates $x^{\cal C}$, which for a given
$x$ take two values $x^-=(x^-_{\mu})$ and $x^+=(x^+_{\mu})$
($\mu\in\{0,1,2,3\}$) on the time ordered and anti-time ordered
branches, respectively (see figure):
%
\begin{eqnarray}
\oint\di x^{\cal C} \dots =\int_{t_0}^{\infty}\di x^-\dots
+\int^{t_0}_{\infty}\di x^+\dots
=\int_{t_0}^{\infty}\di x^-\dots -\int_{t_0}^{\infty}\di x^+\dots,
\end{eqnarray}
%
where only the time limits are explicitly given. One-point
functions  have the same value on both sides on the contour.
The corresponding properties of two-point functions and their
equilibrium relations are summarized in Appendix \ref{discontinouity}. For any
two-point function $F$ the contour values are defined as
$F^{ij}(x,y):=F(x^i,y^j), \quad i,j\in\{-,+\}$ on the different
branches of the contour, cf. (\ref{Fxy}). Due to the change of
operator ordering genuine multi-point functions are discontinuous
in general, when two contour coordinates become identical.

Boson fields may take non-vanishing expectation values of the
field operators $\Pba (x)=\left<\Pa \right>$, called mean fields,
or classical fields. The corresponding equations of motion are
provided by the ensemble average of the operator equations of
motion (\ref{eqmotion})
%
\begin{eqnarray}
\label{eqmotion1}
S_{x}\Pba (x)= - J (x) ,\quad{\rm or}\quad
\Pba (x)=\Pba^0(x) -\oint\di y\Gr^{0}(x,y) J (y).
\end{eqnarray}
%
Here $J (x)=\left<\j (x)\right>$, while $\Pba^0 (x)=\left<\Pa_{\scr{I}}
  (x)\right>$ is the freely evolving classical field, which starts from $\Pba
(t_0,{\vec x})$ at time $t_0$. The free contour
Green's function
%
\begin{eqnarray}\label{G0-int}
\ii\Ga^{0} (x,y&=& \left<\Tc\Ph_{\scr{I}}
(x)\Ph_{\scr{I}}^{\dagger} (y)\right> -\Pba^0(x)(\Pba^0(y))^*
\quad\mbox{is  resolvent of}\\ \label{G0-eq.} S_{x}\Gr^{0}(x,y)
&=&\dc(x,y), \quad\mbox{with}\quad\dc (x^i,y^j) =
\sigma^{ij}\delta^4(x-y)=
\left(\begin{array}{cc}1&0\\0&-1\end{array}\right)\delta^4(x-y).\nonumber
\end{eqnarray}
%
One can easily verify the equivalence of the contour form
(\ref{eqmotion1}) with  the standard retarded classical field
equation and the fact that $J (x)$ and $\Pba (x)$ are one-point
functions, which have identical values on both sides of the
contour.

Subtracting the classical fields via
$\Ph =\Pba +\Pt$,
the full propagator in terms of quantum-fluctuating parts
$\Pt$ of the fields is defined as
%
\begin{eqnarray}\label{Ga1}
\ii\Gr(x,y) = \left<\Tc\Pt (x)\Ptd(y)\right> =\left<\Tc\Ph (x)\Phd
(y)\right> -\phi (x)\phi^{*} (y) =\left<\Tc\Ph (x)\Phd
(y)\right>_c \,.
\end{eqnarray}
%
Index $"c"$ indicates that uncorrelated parts are subtracted. In
terms of diagrams it implies, that the corresponding expectation
values are given by sums of  {\em connected} diagrams.

 Averaging the operator equations of motion (\ref{eqmotion})
multiplied by $\Pad(x)$ and  subtracting classical field parts one
obtains the equation of motion for the propagators,
%
\begin{eqnarray}\label{Dyson}
S_{x}\Gr (x,y) &=& \dc(x,y)+\oint \di z \Se (x,z)\Gr (z,y),
\end{eqnarray}
%
and similarly for the corresponding adjoint equation.
 $\Sa$ denotes the
 proper self-energy of the particle. Since we have separated the full propagator
 in (\ref{Ga1}), $-\ii\Sa$ has to be {\em
  one-particle irreducible} (label ${\rm 1PI}$), i.e. the corresponding
diagram cannot be split into two pieces, which separate $x$ from
$z$ by cutting a single propagator line.  Obviously, $\Se$
can have singular ($\delta$-functional)
one-point parts and genuine two-point parts (the latter are given
by all  connected  1PI
  diagrams of the current--current correlator),
%
\begin{eqnarray}
\label{Se-definition}
-\ii\Sa(x,z)
=
\left<\Tc\frac{\partial^2\ii\Lint(x)}{\partial\Ph
\partial\Phd}\right>_{c} \dc(x,y) -\left<\Tc\j
(x)\j^{\dagger}(y)\right>_{\rm 1PI},
\end{eqnarray}
%
in the Heisenberg picture.

In diagrams free and full classical fields are represented by
''pins'' with cross and ''o-cross'' as heads, cf.
(\ref{Cl-eq-diag}), while free and full propagators are given by
thin and thick long lines, respectively.  Thereby, complex fields
carry a sense, the arrow always pointing towards the $\Ph$ in the
contour ordered expressions. In diagrammatic representation, the
classical field equations (\ref{eqmotion1}) and Dyson's equation
(\ref{Dyson}) are then given by
%
\begin{eqnarray}\label{Cl-eq-diag}
\GPbfull&=&\GPb + \GGJ\vphantom{\displaystyle\int_A^B},\\
\label{Dyson-eq-diag} \GGfull&=&\GG +
\Dysonself\vphantom{\displaystyle\int_A^B}\,,
\end{eqnarray}
%
with the one- and two-point functions $\ii J(x)$ and $-\ii\Se(x,y)$,
as driving terms.

\subsection{Generating Functional $\Gamma$ and $\Phi$}\label{sect-W-Phi}

In \cite{IKV} we have constructed the generating real-time
$\Gamma$-functional in the form
%
\begin{eqnarray}\label{Gammafull}
\displaystyle &&\Gamma\{\phi,\phi^*,\Gr,\lambda \}
=\displaystyle \Gamma^0+\displaystyle
 \oint \di x \Lg^0\{\phi,\partial_\mu\phi\}
\nonumber
\\
&&+ \ii \kappa \left[ \ln\left(1-\odot\Ga^{0}\odot\Sa\right)
+\odot\Ga\odot\Sa \right] +\displaystyle
\Phi\left\{\phi,\phi^*,\Gr,\lambda \right\}, \label{PHIfunct}
\end{eqnarray}
%
defining the auxiliary functional
$\Phi\left\{\phi,\phi^*,\Gr,\lambda \right\}$. To construct this
functional we introduced a space-time dependent interaction scale
$\lambda(x)$ into the variational concept, which scales
interaction vertices, i.e.
%
\begin{equation}
\label{L_l} \Lint_{\lambda}= \lambda(x)\Lint \left\{ \Phd (x),\Ph
(x)\right\}, \quad  {\cal
  E}^{\scr{int}}_\lambda(x)=-\left<\Lint_{\lambda}(x)\right>.
\end{equation}
%
  The
  $\Gamma^0$ and $\Lg^0$ parts, where $\Lg^0$ is the free {\em
classical} Lagrangian function, represent the non-interacting
parts of $\Gamma$.  Thereby,  the $\Gamma^0$ term solely depends
on the unperturbed propagator $\Gr^0$ and hence is treated as a
constant with respect to functional variations of $\Gamma$.  The
$\ln(\dots)$ is understood in the functional sense, i.e. by a
series of $n$-folded contour convolutions, denoted by the
$\odot$-symbol, formally resulting from the Taylor expansion of
the $\ln(1+x)$ at $x=0$. The $\ln$-term accounts for the change of
$\Gamma$ due to the self-energies of the particles. These first
three terms account for the one-body components in the $\Gamma$.
The remaining $\odot\Ga\odot\Sa$ and $\Phi$ terms correct for the
true interaction energy part of the partition sum.

Functional variation of
$\Gamma\left\{\phi,\phi^*,\Gr,\lambda\right\}$ in the form of eq.
(\ref{PHIfunct}) leads to
%
\begin{eqnarray}
\label{vargamma}\hspace*{-1cm} \displaystyle \delta\Gamma
\left\{\phi,\phi^*,\Gr,\lambda\right\} &=&\displaystyle
\kappa\left\{ \oint\di x \left[\delta\Pba (x)(\So_x )^{\ast}
\Pba^*(x)+\delta\Pba^*(x)\So_x \Pba(x)\right]\right. \nonumber
\\
\hspace*{-1cm}&&\hspace*{2cm}-
\ii  \left(
\odot \frac{1}{1-\odot\Ga^{0}\odot\Sa}\odot\Ga^{0} - \odot\Ga
\right)\odot\delta\Sa
\nonumber
\\
\hspace*{-1cm}&&\hspace*{2cm}+\displaystyle\left.
 \ii  \oint \di x \di y \Sa(x,y) \delta\Ga(y,x) \right\}
+\delta\Phi\left\{\phi,\phi^*,\Gr,\lambda\right\}.
\end{eqnarray}
%
Here $\delta\Sa$ is understood as a variation induced by $\delta\Ga$,
$\delta\Pba$, $\delta\Pba^*$, and $\delta\lambda$, respectively.
%
 In terms of $\Gamma\left\{\phi,\phi^*,\Gr,\lambda\right\}$
equations of motion (\ref{eqmotion1}) and (\ref{Dyson}) read
%
\begin{eqnarray}
\label{varG/phi} \delta \Gamma / \delta \Pba = 0, \,\,\, \delta
\Gamma / \delta \Pba^* = 0, \,\,\, \delta \Gamma / \delta \Ga =
0\,,
\end{eqnarray}
%
i.e. the functional variations of $\Gamma$ with respect to $\Ga$,
$\Pba$ and $\Pba^*$ at $\lambda(x) = 1$ vanish for the physical
solutions.
These imply the  following variational rules
for the auxiliary $\Phi$-functional:
%
\begin{eqnarray}
\label{varphi'}
\delta\Phi\left\{\phi,\phi^*,\Gr,\lambda\right\}&=&\displaystyle
\kappa\left\{ \oint\di x
\left[\jba^*(x)\delta\Pba(x)+\jba(x)\delta\Pba^*(x)\right]
\nonumber\right. \\ &&\hspace*{1cm}-\displaystyle\left.
 \ii  \oint \di x \di y
\Sa(x,y)\delta\Ga(y,x)\right\}
-\displaystyle
\oint\di x{\cal E}^{\scr{int}}(x)\delta\lambda(x),
\end{eqnarray}
%
or
%
\begin{eqnarray}\label{varphdl}
\ii\jba(x)&=&\frac{\delta\ii \Phi}{\delta \Pba^{\ast} (x)},\quad
\label{varphdl1}
-\ii \Sa(x,y)=\frac{\delta\ii \Phi}{\delta \ii\Ga(y,x)}\times
\left\{
\begin{array}{ll}
2\quad&\mbox{for real fields}\\[1.5mm]
1\quad&\mbox{for complex fields}
\end{array}\right.\\
\label{varphdl2}
- {\cal E}^{\scr{int}}(x)&=&
\frac{\delta\ii \Phi}{\delta \ii\lambda (x)}.
\end{eqnarray}
%
Thus, $\Phi$ is a {\em generating} functional for the source terms
$\jba$ of classical fields and self-energies $\Sa$. Therefore,
approximation schemes can be defined through a particular
approximation to $\Phi$. The invariance properties of $\Phi$ play
a central role to define conservation laws for the approximate
dynamics.

It is important to emphasize that we do all functional variations
independently on any place of the contour. Thus, different contour
times are considered as independent even though they may refer to
the same physical time. The fact that components of $\Ga$ on the
different branches of the contour are not independent (cf.
(\ref{Fretarded})) for the physical solution, has no importance
for the variational procedure. The reason is that rules
(\ref{Fretarded}) only apply to the physical $\Ga$ and $\Pba$,
which are provided by the stationary ``points'' of the variational
principle, i.e. solving the equations of motion (\ref{eqmotion1}),
(\ref{Dyson}).
Moreover, for the closed
real-time contour the values of  $\Gamma$ and $\Phi$ trivially
vanish, i.e. $\Gamma=\Phi=0$  for physical values of $\Ga$,
$\Pba$,
    $\Pba^*$ and $\lambda=1$.

\subsection{Diagrams for $\Gamma$, $\Phi$ and ${\cal
E}^{\scr{int}}_{\lambda}(x)$}\label{Diagrams}

 According to  (\ref{vargamma}) and (\ref{varphdl2})
%
\begin{eqnarray}\label{ep-phi}
-\oint\di x {\cal E}^{\scr{int}}(x)=
\left[\lambda\frac{\di}{\di \lambda}
\Gamma\{\phi\{\lambda\},\phi^*\{\lambda\},\Gr\{\lambda\},\lambda\}
     \right]_{\lambda=1}
=\left[\lambda\frac{\partial}{\partial\lambda}
\Phi\{\phi,\phi^*,\Gr,\lambda\}\right]_{\lambda=1},
\end{eqnarray}
%
where now $\lambda$ is considered as a {\em global} scale
parameter (to $\Phi$ only a partial derivative is applied, i.e.
the $\Pba$, $\Pba^*$ and $\Ga$ values are kept constants). The
expression
for $-\ii{\cal E}^{\scr{int}}(x)$ can be re-summed and entirely
expressed in terms of full classical fields and full propagators.
The re-summed diagrams are then void of any self-energy insertions
and therefore have to be {\em two-particle irreducible}

 \unitlength=.8cm

%
\begin{eqnarray}\label{Eint-diag1}
-\ii{\cal E}^{\scr{int}}(x) =\sum_{n_\lambda}\Dclosedone{\rm 2PI
}{\thicklines},\vhight{.8}
\end{eqnarray}
%
 i.e. they cannot be
decomposed into two pieces by cutting two propagator lines. The
formal integration of the last equality in (\ref{ep-phi}) with
respect to $\lambda$ keeping $\phi$ and $\Gr$ constant provides
the diagrammatic expression for $\Phi$ in terms of full Green's
functions and classical fields. Therefore
$\ii\Gamma\left\{\phi,\phi^*,\Gr,\lambda\right\}$ can be expressed
in terms of  diagrams (cf. eq.  (\ref{PHIfunct})) as
%
\begin{eqnarray}\label{keediag}
&&\ii\Gamma\left\{\phi,\phi^*,\Gr,\lambda\right\} = \ii
\Gamma^0\left\{\Gr^0\right\}
+ \ii \oint \di x \Lg^0\{\phi,\partial_\mu\phi\}
\nonumber
\\
&&\hspace*{5mm} +\kappa \left\{\vhight{1.6}\right.
\underbrace{\sum_{n_\Se}\vhight{1.6}\frac{1}{n_\Se}\GlnG0Sa}
_{\displaystyle -\ln\left(1-\odot\Ga^{0}\odot\Sa\right)}
\underbrace{-\vhight{1.6}\GGaSa} _{\displaystyle
-\odot\Ga\odot\Sa\vphantom{\left(\Ga^{0}\right)}}
\left.\vhight{1.6}\right\}
\underbrace{+\vhight{1.6}\sum_{n_\lambda}\frac{1}{n_\lambda}
\Dclosed{\rm 2PI}{\thicklines}}
_{\displaystyle\vphantom{\left(\Ga^{0}\right)} +\ii\Phi}.
\end{eqnarray}
%
Here $n_\Se$ counts the number of $\Sa$ insertions in the ring diagrams
providing the $\ln$-terms, while for the closed diagrams of $\Phi$ the value
$n_\lambda$ counts the number of vertices building up the functional $\Phi$.
The diagrams contributing to $\Phi$ are given in terms of full
propagators $\Ga$ and full time-dependent classical fields $\Pba$.
As a consequence, these diagrams have to be two-particle
  irreducible.  The latter property is required because of the
re-summations of ${\cal E}^{\scr{int}}(x)$.  This also matches
diagrammatic rules for the re-summed self-energy $\Sa(x,y)$, which
results from functional variation of $\Phi$ with respect to any
propagator $\Ga(y,x)$. In graphical terms, this variation is
realized by opening a propagator line in all diagrams of $\Phi$.
The resulting set of thus opened diagrams must  be that of proper
skeleton diagrams of $\Sa$ in terms of {\em full propagators},
i.e. void of any self-energy insertion.

The diagrammatic rules for $\Phi$, ${\cal E}^{\scr{int}}(x)$, $\jba$
and $\Sa$ are as in perturbation theory, except that
(i) for all bosonic fields in $\ii\Lint$, replace $\Pa$ by
$\Pba+\Pta$ in order to account for the classical fields;
(ii) all pair contractions represent full propagators
 $\ii\Ga(x,y)$;
(iii) keep only those diagrams  that correspond to two-particle
irreducible diagrams for $\Phi$, i.e. which cannot be split into
two pieces by cutting two different propagator lines.
 For  the rules in the $\{-+\}$ matrix notation we refer to
\cite{Lif81}.

 As an example, we quote
the diagrams in neutral scalar $g \Ph^4/4!$ theory. Up to two
vertices, the functional $\Phi$ is given by the following
expressions
%
\begin{eqnarray}
\label{Phi/phi4}
\ii \Phi&=&\frac{-\ii g}{4!} \oint\di x
\left(\phi^4(x)+6\phi^2(x)
\left<\Pta(x)\Pta(x)\right>_c+
3\left<\Pta(x)\Pta(x)\right>_c^2\right)
\nonumber
\\[-2mm]&&\\[-2mm]
&+&
\frac {1}{2} \left(\frac{-\ii g}{4!}\right)^2
\oint\di x\oint\di y \left(4\cdot4!\phi(x)\phi(y)
\left<\Pta(x)\Pta(y)\right>_c^3
+{4!}\left<\Pta(x)\Pta(y)\right>_c^4\right)+\dots,
\nonumber
\end{eqnarray}
%
or in terms of diagrams
%
\unitlength=.600mm
\begin{eqnarray}\label{phi4diagrams}\linethickness{3.pt}
\begin{array}{lccccccccccc}
\ii\Phi &=&  \gphifour
&+&  \gloopphitwo
&+& \geight
&+\displaystyle\frac{1}{2}\left\{\vphantom{\rule{0mm}{1.cm}}\right.&
        \gphisand3phi
&+&\gsand4
&\left.\vphantom{\rule{0mm}{1.cm}}\right\}
        +\displaystyle\frac{1}{3}\dots\vphantom{\rule{0mm}{1.cm}}
\\[7mm]
&&\left[\frac{1}{4!}\right]
&&\left[\frac{1}{2\cdot2!}\right]
&&\left[\frac{1}{2^2\cdot 2!}\right]
&&\left[\frac{1}{3!}\right]
&&\left[\frac{1}{4!}\right]&
\end{array}\nonumber\\[-0.6cm]
\end{eqnarray}
%
The $1/n_\lambda$ factors are explicitly given, while the standard
combinatorial factors are given in square brackets below each
diagram. Functional derivatives with respect to $\phi$ (pins) and
propagators (full lines),  cf. eqs. (\ref{varphi'}),
(\ref{varphdl}), determine the source $J(x)$ of the classical
field and the self-energy $\Se(x,y)$, respectively,
%
\begin{eqnarray}
\label{J/phi4}
\begin{array}{rccccccccl}
\ii J(x)
&=&\gphithree &+& \gloopphione &+&
\gloopphih\;\;&+&\dots\vphantom{\rule{0mm}{1.cm}} ,\\[7mm]
&&\left[\frac{1}{3!}\right]
&&\left[\frac{1}{2}\right]
&&\left[\frac{1}{3!}\right]
&\\[6mm]
- \ii \Se(x,y)
&=& \sphitwo &+&
 \sloop &+&
\sphisand2phi
&+&\ssand3&+\dots\vphantom{\rule{0mm}{1.cm}}\\[7mm]
&&\left[\frac{1}{2!}\right]
&&\left[\frac{1}{2}\right]
&&\left[\frac{1}{2!}\right]
&&\left[\frac{1}{3!}\right]
&
\end{array}
\end{eqnarray}
%
Small full dots define vertices, which are to be integrated over,
while big full dots specify the external points $x$ or $y$; the
first two diagrams of $\Se(x,y)$ give the singular $\dc(x,y)$
parts arizing from classical fields and tad-poles.

\subsection{$\Phi$-Derivable Approximations and Invariances of
$\Phi$ }

The expressions for  $\Gamma$ and $\Phi$ given above are exact and
expressed  in terms of full propagators and self-energies. The
$\Phi$-derivable approximations \cite{Baym} are constructed by
confining the infinite diagrammatic series for $\Phi$ either to a
set of a few diagrams or some sub-series of diagrams.  Thereby the
approximate $\Phi^{\scr{(appr.)}}$  is constructed in terms of
{\em full} Green's functions and {\em full} classical fields,
where {\em full} now takes the sense of solving self-consistently
the classical field and Dyson's equation with the driving terms
derived from $\Phi^{\scr{(appr.)}}$ through relations
(\ref{varphdl}). It means that even restricting ourselves to a
single diagram in $\Phi^{\scr{(appr.)}}$, in fact, we deal with a
whole sub-series of perturbative diagrams. $\Phi^{\scr{(appr.)}}$
serves as a generating functional for the approximate source
currents $\jba^{\scr{(appr.)}}(x)$ and self-energies
$\Sa^{\scr{(appr.)}}(x,y)$ (see  eqs. (\ref{varphdl}) )
%
\begin{equation}\label{varphdl/appr}
\ii\jba^{\scr{(appr.)}}(x)
=\frac{\delta\ii \Phi^{\scr{(appr.)}}}{\delta
\left(\Pba^{\scr{(appr.)}\ast} (x)\right)},
\end{equation}
\begin{equation}\label{varphdl1/appr}
-\ii
\Sa^{\scr{(appr.)}}(x,y)=
\frac{\delta\ii \Phi^{\scr{(appr.)}}}{\delta \ii\Ga^{\scr{(appr.)}}(y,x)}
\times\left\{\begin{array}{ll}
2\quad&\mbox{for neutral fields}\\
1\quad&\mbox{for charged fields}
\end{array}\right.
\end{equation}
%
which then are the driving terms for the equations of motion for
the classical fields and propagators. $\Phi^{\scr{(appr.)}}$ also
provides the corresponding expression for ${\cal E}^{\scr{int}}$
(see eq. (\ref{varphdl2})). Below, we omit the superscript
``appr.''.

The invariances of $\Phi$ play as central  role as the invariances
of the Lagrangian for the full theory.  Thereby, the variational
principle, where the interaction strength $\lambda(x)$, the
classical fields $\Pba(x)$ and propagators $\Ga(x,y)$ can be
varied independently, provides a set of useful identities and
relations.  A general invariance of $\Phi$ is provided by the
substitution $x\Rightarrow x+\xi(x)$ for all integration variables
in the contour integrations defining $\Phi$. This invariance
results in the energy-momentum conservation,
$\partial_\mu \Theta^{\mu\nu}(x)=0$, with the
energy-momentum tensor given by the Noether expression
%
\begin{eqnarray}
\label{E-M-G-tensor} \nonumber \Theta^{\mu\nu}
 &=&
\frac{1}{2}\kappa \left[ \left(\left(\partial^{\nu}_x\right)^* +
\partial^{\nu}_y\right) \left(\left(\partial^{\mu}_x\right)^* +
\partial^{\mu}_y\right) \left(\vphantom{\frac{1}{2}} \Pba^*(x)
\Pba(y) +
\ii\Ga (y,x)\vphantom{\frac{1}{2}}\right)
\right]_{x=y}
\nonumber
\\
&&+
g^{\mu\nu}
\left({\cal E}^{\scr{int}}(x)-
{\cal E}^{\scr{pot}}(x)
\right)
\end{eqnarray}
%
with the interaction energy density (\ref{varphdl2}) and ${\cal E}^{\scr{pot}}$
expressed as
%
\begin{eqnarray}
\label{eps-pot1} {\cal E}^{\scr{pot}}(x)
&=&\displaystyle\frac{1}{2}\kappa \left\{ \vphantom{\oint}
-
\left[\jba^*(x)\Pba(x) + \jba(x)\Pba^*(x)\right]
\right. \nonumber
\\
&+&\displaystyle\left. \ii
\oint\di
z\left[\Sa(x,z)\Ga(z,x)+\Ga(x,z)\Sa(z,x)\right]
\right\}.
\end{eqnarray}
%

Along similar lines charge conservation can be proven, provided
$\Phi$ is invariant under  the simultaneous variation of
classical fields and propagators
%
\begin{equation}
\label{c-global-tr1}
\Pba(x)\Rightarrow e^{-\ii e\Lambda (x)}\Pba(x),\quad
\Pba^*(x)\Rightarrow e^{\ii e\Lambda (x)}\Pba^*(x),\quad
\Ga(x,y)\Rightarrow e^{-\ii e\Lambda (x)}
\Ga(x,y)e^{\ii e\Lambda (y)}.
\end{equation}
By means of equations of motion (\ref{eqmotion1}), (\ref{Dyson})
the divergence of the  Noether current
%
\begin{eqnarray}
\label{c-G-current} j^{\mu} (x) &=& e \left[
\left(\left(\partial^{\mu}_x\right)^* + \partial^{\mu}_y\right)
\left(\vphantom{\frac{1}{2}} \Pba^*(x) \Pba(y) +
\ii\Ga (y,x)\right) \right]_{x=y}
\end{eqnarray}
%
 vanishes.
Thus,
the current conservation takes place for any $\Phi$-derivable approximation,
which is invariant with respect to (\ref{c-global-tr1}).

Further invariances generally depend on the properties of the
interaction vertices in the theory considered. An example is the
invariance discussed in the context of eq. (\ref{gamma}), which
now transcribes to the corresponding expectation values.

\subsection{Generating Functional $\Gamma^{\rm eq}$ and Thermodynamic Consistency} \label{Consistency}

In the thermal equilibrium the thermodynamic potential is
explicitly known, cf. \cite{Abrikos},
%
\begin{eqnarray}
\label{om-eq-Z}
\Omega  = - T \ln Z,
\quad\mbox{where}\quad
\label{ro-eq}
\widehat{\rho}^{\scr{eq}} =
\frac{\exp\left(-\beta\Hh\{\mu\}\right)}{Z},
\quad\mbox{with}\quad
\Tr\widehat{\rho}^{\scr{eq}} = 1,
\end{eqnarray}
%
where $\beta = 1/T$ is the inverse temperature,
$\widehat{\rho}^{\scr{eq}}$ is the  equilibrium density operator
and $Z$ is the partition function,
%
\begin{eqnarray}
\label{H-mu} \Hh (\mu) = \Hh - \int \di^3 x \mu
\medhat{j}^{0}_{\scr{(Noether)}}(x),
\end{eqnarray}
%
where $\medhat{j}^{0}_{\scr{(Noether)}}$ is the time-component of
the charge current, $\mu$ is the chemical potential.

The equilibrium density matrix formally coincides with evolution
operator in imaginary time, i.e. one writes $\Tr
\widehat{\rho}^{\scr{eq}}...$ instead of $\left<...\right>$. Thus,
 we arrive at the following form of $\Gamma$
functional in equilibrium:
%
\begin{eqnarray}
\label{W-eq} \hspace*{-5mm}
\Gamma^{\scr{eq}}\left\{\phi,\phi^*,\Gr,\lambda,\mu \right\}
=-\ii\ln\left(\frac{1}{Z} \Tr \exp\left[-\ii\oint\di t
\Hh_{\scr{I}}^{0}\{\mu\} \right] \Tc\exp\left[ \ii\oint\di x
\lambda\Lint_{\scr{I}}\{\mu\}
   \right]\right),
\end{eqnarray}
%
with the integration contour ${\cal C}$ now being the sum of the
real-time Schwinger-Keldysh contour (see figure) and the
imaginary-time Matsubara contour ${\cal C_{\scr{eq}}}$, i.e. it
starts from an initial  time $t_0$, goes to infinity, then back to
this initial time and after that, to $t_0-\ii\beta$. Taking into
account the fact that $\Gamma = 0$ for the physical values of
$\phi$, $\phi^*$, $\Gr$, and $x$-independent $\lambda$, from eq.
(\ref{om-eq-Z}) we obtain
%
\newcommand{\eqoint}{\int_{\cal C_{\scr{eq}}}}
\begin{eqnarray}
\label{Om-eq} \hspace*{-3mm} \Omega
\{\phi,\phi^*,\Gr,\lambda,\mu\} = - T \ln \left\{\Tr \left(
\exp\left[-\ii\eqoint\di t \Hh_{\scr{I}}^{0}\{\mu\} \right]
\Tc\exp\left[ \ii\eqoint\di x \lambda\Lint_{\scr{I}}\{\mu\}
   \right]\right)\right\}.
\end{eqnarray}
%
where the integral over the real-time contour section gives zero.
In eq. (\ref{Om-eq}) we can make the replacement
%
\begin{eqnarray}
\label{C_FS->C_M} \eqoint\di t ... = \int_0^{-\ii \beta} \di t ...
\quad
\end{eqnarray}
%
Thus, one arrives at the proper representation of the
thermodynamic potential originally proposed by Luttinger and Ward
\cite{Luttinger}. Indeed, since all quantities under the integral
are analytically continued from the Schwinger--Keldysh contour to
the Matsubara contour, $\Omega$ is determined by the same
expression as the $\Gamma$-functional (\ref{PHIfunct}) but in
terms of the Matsubara Green's functions with the thermodynamic
$\Phi_T$-functional represented by the same set of closed
diagrams.
Thus, the
problem of the thermodynamic consistency is re-addressed from the
Schwinger--Keldysh approach to the Matsubara one. Within the
Matsubara formalism, ref. \cite{Baym} has shown that any
$\Phi$-derivable approximation to the thermodynamic potential is
thermodynamically consistent. Hence, the $\Phi$-derivable
approximations to  non-equilibrium $\Gamma$-functional are also
thermodynamically consistent \cite{IKV}.

The stationary property of the $\Gamma$ functional (and, hence, of
$\Omega$) with respect to variations in full Green's functions and
classical fields, eq. (\ref{varG/phi}), is the key feature of
$\Omega$ that provides the thermodynamic consistency. It implies
that any derivative of the partition sum to any thermodynamic
parameter like $\beta$ or $\mu$ is then given by accounting only
the explicit dependence of $\Omega$ on these parameters, since the
implicit dependences through $\Ga$ and $\phi$ drop out due to the
stationarity. Therefore $\Phi$-derivable approximations preserve
the corresponding properties of the exact theory, providing
thermodynamic consistency.


\section{Non-equilibrium and Thermodynamic Potentials}
\label{Potential}

It is instructive to introduce a non-equilibrium potential
$\widetilde{\Omega}$ and to get the relation between this
potential, the $\Gamma$-functional introduced in eq.
(\ref{PHIfunct}), and the thermodynamical potential $\Omega$ (in
thermal equilibrium). This relation will provide another and more
direct way for demonstrating thermodynamic consistency of
$\Phi$-derivable approximations in the spirit of Baym \cite{Baym}
 and also can be useful in certain applications of the
$\Phi$-derivable approach.

For the sake of convenience, we introduce the zero component of
the operator  $\widehat{p}_x^0 $  shifted by the chemical
potential
%
\begin{eqnarray}
\label{pi0-operator} \widehat{\pi}_x^0 =  \ii (\partial_t - \ii e
\mu, -\nabla_x) \,.
\end{eqnarray}
%
We have introduced the chemical potential in such a way that it
fixes a conserving quantity, related to some charge  (e.g.,
electric charge,  strangeness, etc.).

Let us introduce an auxiliary non-equilibrium potential
$\widetilde{\Omega}$:
%
\begin{equation}
\label{Om-Gamma} \Gamma\left\{\phi,\phi^*,\Gr,\lambda,\mu\right\}
= - \oint \di t
\widetilde{\Omega}\left\{\phi,\phi^*,\Gr,\lambda,\mu\right\}.
\end{equation}
%
Unlike $\Gamma$, the new quantity $\widetilde{\Omega}$ is non-zero
for physical values of Green's functions and mean fields. The
value $-\widetilde{\Omega}$ plays the role of the effective
Lagrangian. Below we show that in thermal equilibrium this
non-equilibrium potential has the meaning of the usual
thermodynamic potential. In order to demonstrate this, we first
find variations of $\widetilde{\Omega}$ with respect to the
chemical potential and to the $\lambda$-vertex scaling parameter
at arbitrary non-equilibrium. Based on the identity
(\ref{Om-Gamma}), we can extend the method of the on-contour
variations of $\Gamma$ in order to get derivatives of
non-equilibrium and thermodynamic potentials.

The $\Gamma$-functional introduced above
is now  a functional of the chemical potential. This functional
dependence can be treated in two ways. First, we can consider the
functional dependence of $\Gamma$ on $\mu$ as originated solely
from changes of Green's functions, self-energies and mean fields,
which are induced by a variation of the chemical potential.
Indeed, the additional term of eq. (\ref{H-mu}) can be completely
absorbed into equations of motion by changing
%
$\partial_t \to \partial_t - \ii e \mu$
%
in $\So_x$ operators of eq. (\ref{eqmotion}).  Notice that now the
free Green's functions (cf. eq. (\ref{G0-eq.})) also depend on the
chemical potential rather than only full Green's functions and
self-energies.

Let us calculate the variation of $\Gamma$ with respect to $\delta
\mu$ in the above-described way, provided all other variables of the
$\Gamma$-functional are kept constant. Here, we allow the
variation of $\mu$ on the contour, i.e., $\delta\mu (x)$ is a
contour function, which, in general, takes different values on
different branches of the contour. To get the variation  $\delta
\Ga^0 (x,y)$ induced by $\delta \mu (x)$, we vary eq.
(\ref{G0-eq.}). Then we obtain
%
\begin{equation}
\label{dG-free/mu-eq} \So_x(\mu) \delta \Ga^0 (x,y) = - \delta
\So_x(\mu) \Ga^0 (x,y),
\end{equation}
%
where
%
\begin{equation}
\label{dS/mu} \delta \So_x(\mu) = - e \left(\delta \mu (x)
\widehat{\pi}^0_x + \widehat{\pi}^0_x \delta \mu (x) \right),
\end{equation}
%
or, by using the fact that $\Ga^0$ is the resolvent of the $\So_x$
operator, we arrive at
%
\begin{eqnarray}
\label{dG-free/mu} \delta \Ga^0 (x,y) &=& - \oint \di z \Ga^0
(x,z) \delta \So_z(\mu) \Ga^0 (z,y) \nonumber \\ &=& - \oint \di z
\delta \mu (z) e \left(  \Ga^0 (x,z) \left[\widehat{\pi}^0_z \Ga^0
(z,y)\right] - \left[\widehat{\pi}^0_z \Ga^0 (x,z)\right]\Ga^0
(z,y) \right).
\end{eqnarray}
%

To calculate variations with respect to $\Ga^0$ we need the rule
to vary $\Gamma^0$:
%
\begin{eqnarray}
\label{Gamma0-var} \delta \Gamma^0 = -\ii
\kappa\oint \di x \left(\So_x \delta \Ga^0(x,y)\right)_{y=x},
\end{eqnarray}
%
which  follows from the definition of $\Gamma^0$
(cf. eq. (\ref{PHIfunct})). Varying now the free part of $\Gamma$
in $\mu$ according to eq. (\ref{Gamma0-var}), we obtain
%
\begin{equation}
\label{G0-var-mu} \delta \Gamma^0 = \ii
\kappa e \oint \di x \left[\left( \vphantom{\frac{1}{m_a}}
\widehat{\pi}^0_x + (\widehat{\pi}^0_y)^* \right) \Ga^0
(x,y)\right]_{y=x} \delta \mu (x).
\end{equation}
%
The interaction part of $\Gamma$ (cf. eq. (\ref{PHIfunct}))
depends on $\mu$ only through $\Ga$, $\Sa$, $\Ga^0$, mean fields
and their derivatives. Note that $\Gamma$ is stationary under
variations in $\Ga$ and mean fields near their physical values.
Hence, the variation of $(\Gamma-\Gamma^0)$ results only from the
variations of $\Ga^0$ and mean-field derivatives in
$\Lg^0\{\phi,\partial_\mu \phi\}$. Thus, varying
$(\Gamma-\Gamma^0)$, we get (in symbolic form of eq.
(\ref{PHIfunct}))
%
\begin{eqnarray}
\label{G-G0-var-mu1} &&\delta \left(\Gamma - \Gamma^0\right) =
\kappa e \oint  \di x\left[\Pba^* \cdot \widehat{\pi}^0_x \Pba +
\left(\widehat{\pi}^0_x\right)^* \Pba^* \cdot \Pba\right]
\delta\mu(x) \nonumber
\\
&&- \ii
\kappa\odot \frac{1}{\left(1-\odot\Ga^{0}\odot\Sa\right)} \odot
\delta\Ga^{0}\odot\Sa\,.
\end{eqnarray}
%
 Substituting here $\delta\Ga^{0}$ in the form of
eq. (\ref{dG-free/mu}), we arrive at
%
\begin{eqnarray}
\label{G-G0-var-mu} \delta \left(\Gamma - \Gamma^0\right) &=&
\kappa e \oint  \di x\left[\Pba^* \cdot \widehat{\pi}^0_x \Pba +
\left(\widehat{\pi}^0_x\right)^* \Pba^* \cdot \Pba\right]
\delta\mu(x) \nonumber
\\
&+&\ii
\kappa e\oint \di x \di z \di z' \left[\left(
\vphantom{\frac{1}{m_a}} \widehat{\pi}^0_x + (\widehat{\pi}^0_y)^*
\right) \Ga^0 (x,z) \Sa(z,z') \Ga (z',y)\right]_{y=x} \delta \mu
(x).
\end{eqnarray}
%
It is important to point out that we have essentially used the
variation rules (\ref{varphi'}) to get this expression. Thus, the
final result is
%
\begin{eqnarray}
\label{G-var-mu} \delta \Gamma &=& \oint \di x \delta \mu (x)
\kappa e \left( \vphantom{\suma} \left[\Pba^* \cdot
\widehat{\pi}^{0}_x \Pba + \left(\widehat{\pi}^{0}_x\right)^*
\Pba^* \cdot \Pba\right]
\right. \nonumber
\\
&+&\ii
\left.\left[\left( \vphantom{\int} \widehat{\pi}^0_x +
(\widehat{\pi}^0_y)^* \right) \Ga (x,y)\right]_{y=x}
\vphantom{\suma}\right)\,.
\end{eqnarray}
%
Comparing the expression under the integral with that for the
current (\ref{c-G-current}), we see that the above variation takes
the form
%
\begin{equation}
\label{var-mu} \delta \Gamma\left\{
\phi\{\mu\},\phi^*\{\mu\},\Gr\{\mu\},\lambda=\mbox{const},
\mu(x)\right\}  = \oint \di x j^0 (x) \delta \mu (x),
\end{equation}
%
where $j^0$ is the zero-component of the current, i.e. the
density. In all we have done so far, we did not assume the thermal
equilibrium. Hence, eq. (\ref{var-mu}) holds for any
non-equilibrium.

If we perform all the same manipulations for the equilibrium
$\Gamma$-functional ($\Gamma^{\rm eq}$) defined by eq.
(\ref{W-eq}), we arrive at precisely the same result
(\ref{var-mu}) but in equilibrium. On the other hand, in
equilibrium we can straightforwardly vary eq. (\ref{W-eq}) in the
chemical potential and, hence,  immediately arrive again at eq.
(\ref{var-mu}).  The coincidence of these two ways of variation
shows that we are able to consistently introduce chemical
potentials only provided we deal with a $\Phi$-derivable
approximation, i.e. when the variation rule (\ref{varphi'}) holds
true.

Now, comparing eq. (\ref{var-mu}) with eq. (\ref{Om-Gamma}), we
obtain 
%
\begin{equation}
\label{O-var-mu} \delta \widetilde{\Omega}\{\mu(x)\} = - \int
\di^3 x j^0 (x) \delta \mu (x) .
\end{equation}
%
This exactly coincides with the respective derivative of the
thermodynamic potential $\Omega$ in equilibrium
%
\begin{equation}
\label{O-var-mu-eq} \frac{\partial \Omega (\mu)}{\partial \mu} = -
\int \di^3 x j^0\equiv -N_e\,,
\end{equation}
%
 thus yielding the total charge $N_e$ associated with $j$-current.  
 Therefore, we conclude that
%
\begin{equation}
\label{varOm-mu} \int \di^3 x\left( \frac{\delta
\widetilde{\Omega} \{\mu\}}{\delta \mu} \right)_{\scr{eq}} =
\frac{\partial \Omega (\mu)}{\partial \mu}.
\end{equation}
%

We can also find a variation of $\widetilde{\Omega}$ with respect to
$\lambda$ being considered a contour function of time. From eqs.
(\ref{vargamma}), (\ref{varphi'}) we have
%
\begin{eqnarray}
\label{varGam-lambda} \delta \Gamma\left\{
\phi\{\lambda\},\phi^*\{\lambda\},\Gr\{\lambda\},\lambda(x),
\mu=\mbox{const}\right\} = - \oint \di t\int \di^3 x {\cal
E}^{\scr{int}}_\lambda (x) \delta \lambda (x)\,,
\end{eqnarray}
%
or in terms of the non-equilibrium potential
%
\begin{equation}
\label{Om-eps-int} \delta \widetilde{\Omega}\{\lambda\} = \int
\di^3 x {\cal E}^{\scr{int}}_\lambda (x) \delta \lambda (x).
\end{equation}
%

On the other hand, in statistical physics, cf. \cite{Abrikos},
there is a well-known theorem relating a variation of the
thermodynamical potential $\Omega$ in a parameter $\lambda$ to the
statistical average of the corresponding variation of $\Lint$ in
the same parameter
%
\begin{equation}
\label{stat.phys.} \frac{\partial \Omega(\lambda)}{\partial
\lambda} = - \int \di^3 x \left< \frac{\partial \Lint}{\partial
\lambda} \right>  = \int \di^3 x {\cal E}^{\scr{int}}_\lambda
(x)\,.
\end{equation}
%
With $\lambda$, being the scaling factor of the interaction
Lagrangian of eq. (\ref{L_l}), this relation follows from
expression  (\ref{Om-eq}) for the thermodynamic potential. It is
completely similar to eq. (\ref{Om-eps-int}) for the
non-equilibrium potential. Thus, we obtain in equilibrium
%
\begin{equation}
\label{varOm-lambda} \int \di^3 x\left( \frac{\delta
\widetilde{\Omega} \{\lambda\}}{\delta \lambda} \right)_{\scr{eq}}
= \frac{\partial \Omega (\lambda)}{\partial \lambda}.
\end{equation}
%
Hence, proceeding from properties (\ref{varOm-mu}) and
(\ref{varOm-lambda}), we conclude that  the above-introduced
non-equilibrium potential has  the meaning of the
thermodynamical potential in equilibrium
%
\begin{equation}
\label{Om-eq-noneq} \widetilde{\Omega}_{\scr{eq}} = \Omega .
\end{equation}
%
Thus defined $\widetilde{\Omega}$ is an extension of the
thermodynamic potential to non-equilibrium processes. In a way,
this non-equilibrium potential is equivalent to the
$\Gamma$-functional, while the former has the advantage that it is
non-zero for physical values of Green's functions and mean fields.

Relation (\ref{Om-Gamma}) between $\Gamma$ and
$\widetilde{\Omega}$ allows us to directly evaluate
$\widetilde{\Omega}$ in terms of transport quantities. In
particular, a  useful scheme of calculating the non-equilibrium and
thermodynamic potentials consists in integrating the interaction
energy over the coupling constant (or $\lambda$, in our case), cf.
\cite{Abrikos} for equilibrium systems. Indeed, from eq.
(\ref{Om-eps-int})  we get
%
\begin{equation}
\label{Om-eps-int0} \widetilde{\Omega} 
= \int \di^3 x
\int_{0}^1 \di\lambda {\cal E}^{\scr{int}}
_{\lambda} (x)+ \widetilde{\Omega}_{0},
\end{equation}
%
where $\widetilde{\Omega}_{0}$ is a $\lambda$-independent part,
corresponding to the system of non-interacting particles.
Analogous expression is obtained from eq. (\ref{stat.phys.}) in
the equilibrium case. For specific interactions with a certain
number $\gamma$ of operators attached to the vertex,
with the help of eq. (\ref{gamma}) we  obtain
%
\begin{equation}
\label{Om-eps-int-gam} \widetilde{\Omega}= 
\frac{2}{\gamma} \int \di^3 x \int_{0}^1 
\di \lambda {\cal E}^{\scr{pot}}_{\lambda} (x)+
\widetilde{\Omega}_{0}.
\end{equation}
%
Applying eq. (\ref{eps-pot1}) 
we arrive at the  expression
%
\begin{eqnarray}
\label{Om-eps-pot1} \widetilde{\Omega} 
&=&
\frac{2}{\gamma} \int \di^3 x \int_{0}^1 \frac{\di
\lambda}{\lambda} \left(
\displaystyle\frac{1}{2}
\kappa\left\{ \vphantom{\oint}
-
\left[ \Pba^*(x)\jba(x)+\jba^*(x)\Pba(x)\right] \right.
\right.\nonumber
\\
&+&\displaystyle\left.\left. \ii
\oint\di z\left[\Sa(x,z)\Ga(z,x)+\Ga(x,z)\Sa(z,x)\right]\right\}
\vphantom{\suma}\right) +\widetilde{\Omega}_{0}.
\end{eqnarray}
%
%
%
which allows us to evaluate it directly in terms of non-equilibrium
transport quantities.

In equilibrium, $\Ga$ and $\Sa$ are given by expressions of eq.
(\ref{Geq}).
Then (\ref{Om-eps-pot1}) transforms into the thermodynamic
potential
%
\begin{eqnarray}
\label{Om-eps-pot-mod} \Omega 
&=& \frac{2}{\gamma} \int
\di^3 x \int_{0}^1 \frac{\di
\lambda}{\lambda}
\kappa\left( -\frac{1}{2} \left[
\Pba^*(x)\jba(x)+\jba^*(x)\Pba(x)\right] \vphantom{\int
\dpi{p}}\right. \nonumber
\\
&+& \left. \int \dpi{p} n(p_0 )
\left[\mbox{Re}\Pi^{R}(\lambda ) A (\lambda )
+\mbox{Re}\Delta^{R}(\lambda ) \Gamma (\lambda
)\right]\right) +\Omega_{0},
\end{eqnarray}
%
where $\A=-2\Im\Ga^R$ and $\Gm=-2\Im\Sa^R$ are the spectral
function and spectral width, respectively, while $n(p_0 )$ is the
thermal Bose-Einstein occupation number (\ref{occup}).

\section{Non-Equilibrium Potential, Stress Tensor,
Pressure and Bulk Viscosity}\label{bulk} \label{Pressure}

Let us consider $\delta\Gamma$ induced by an infinitesimal scaling
transformation of the space on the contour
%
\begin{equation}
\label{x->x'} \vec{x}' = (1 + \xi(t)) \vec{x}, \quad \xi(t) \ll 1.
\end{equation}
%
This transformation results in a change of the volume of the
system
%
\begin{equation}
\label{V->V'} V' = \int \di^3 x' = (1 + \xi(t))^3 \int \di^3 x
\simeq (1 + 3 \xi(t)) V.
\end{equation}
%
Under this transformation the $\So_x$ operator changes as
%
\begin{equation}
\label{dS/V} \delta \So_x(\mu) = 2 \xi(t)
\widehat{\vec{p}}_x^2,
\end{equation}
%
where $\widehat{\vec{p}}_x$ is the spatial vector related to
$\widehat{p}^\mu_x = -\ii \partial^\mu$. The variation
of the free Green's function is determined by the equation
%
\begin{equation}
\label{dG-free/V-eq} \So_x \delta \Ga^0 (x,y) = - \delta \So_x
\Ga^0 (x,y) - 3\xi(t) \dc(x,y),
\end{equation}
%
where we have taken into account that the scaling transformation
(\ref{x->x'}) also modifies the contour $\delta$-function. Using
the fact that $\Ga^0$ is the resolvent of the $\So_x$ operator, we
get
%
\begin{eqnarray}
\label{dG-free/V} \delta \Ga^0 (x,y) = - \oint \di z \Ga^0 (x,z)
\delta \So_z(\mu) \Ga^0 (z,y) - 3\xi(t)\Ga^0 (x,y).
\end{eqnarray}
%
Similarly, the scale transformation (\ref{x->x'}) induces
variations $\delta \Ga$, $\delta \Sa$, $\delta \phi$ and $\delta
\phi^*$. However, in view of the stationary properties of the
$\Gamma$-functional, we do not need explicit forms of these
variations. Thus, variation of the $\Gamma$ is determined by the
variation of $\Ga^0$ and variations of internal integrations
entering into the definition of the $\Gamma$-functional, eq.
(\ref{PHIfunct}). For the $\Phi$-functional the variations of
internal integrations can be associated with variation of
$\lambda$
%
\begin{equation}
\label{V->lambda} \int \di^3 x' ...  = \int \di^3 x (1 + \xi(t))^3
... = \int \di^3 x \lambda' ...,
\end{equation}
%
where
%
\begin{equation}
\label{d-lambda/V} \delta \lambda  = \lambda' - 1 \simeq 3\xi(t) .
\end{equation}
%
Now performing the variation of $\Gamma^0$, according to eqs.
(\ref{Gamma0-var}), (\ref{dG-free/V}) and (\ref{dS/V}) we obtain
%
\begin{equation}
\label{G0-var-V} \delta \Gamma^0 = \ii
\kappa \oint \di x \left[\frac{1}{6}
\left( \widehat{\vec{p}}_x + (\widehat{\vec{p}}_y)^* \right)^2
\Ga^0 (x,y)\right]_{y=x} 3 \xi (t).
\end{equation}
%
Note that the contribution  $- 3\xi(t)\Ga^0$ in the $\Ga^0$
variation of eq. (\ref{dG-free/V}) has exactly cancelled out the
contribution emerged from the variation of the internal
integration. Let us now turn to the mean-field term, i.e. the
second term in  (\ref{PHIfunct}). The variation of this term is
induced by the changes of internal integration and mean-field
gradients in  $\Lg^0$
%
\begin{equation}
\label{grad-var-V} \frac{\partial}{\partial \vec{x}'} \simeq (1 -
\xi(t)) \frac{\partial}{\partial \vec{x}},
\end{equation}
%
as well as by variations of mean fields
%
\begin{eqnarray}
\label{Lo-var-V} &&\delta \left( \oint \di x \Lg^0 \{\phi,
\partial_\mu\phi\} \right) = -
\kappa \oint \di x \frac{1}{2} \left(\Pba^* \jba + \jba^* \Pba
\right) 3\xi(t) \nonumber \\ &-&
\kappa \oint \di x \frac{1}{6} \left[\left(\nabla_x - \nabla_y
\right)^2 \Pba^*(x) \Pba(y)\right]_{y=x} 3\xi(t) +
O\left(\delta\Pba,\delta\Pba^* \right).
\end{eqnarray}
%
Here the first term on the r.h.s. results from the variation of
the internal integration, the second, from the variation of
gradient terms, while $O\left(\delta\Pba,\delta\Pba^* \right)$
denotes the contribution of variations of mean fields. We do not
present this last term in an explicit form, since any case it is
cancelled out by the respective variations of the
$\Phi$-functional. This is the stationary property of the
$\Gamma$-functional with respect to variations of mean fields near
their physical values.

Now consider the main interaction terms in the
$\Gamma$-functional. We obtain
%
\begin{eqnarray}
\label{(G-G0-Lo)-var-V} &&\delta \left( \Gamma - \Gamma^0 - \oint
\di x \Lg^0 \{\phi, \partial_\mu\phi\} \right) \nonumber \\ &=&
\ii
\kappa \oint \di x \di z \di z' \left[\frac{1}{6}
\left( \widehat{\vec{p}}_x + (\widehat{\vec{p}}_y)^* \right)^2
\Ga^0 (x,z) \Sa (z,z') \Ga (z',y)\right]_{y=x} 3 \xi (t) \nonumber
\\ &+& \ii \frac{1}{2}
\kappa \oint \di x \di z \left[ \Ga (x,z) \Sa (z,x) + \Sa (x,z)
\Ga (z,x)\right] 3 \xi (t) \nonumber \\ &-& \oint \di x {\cal
E}^{\mbox{\scriptsize int}}(x) 3 \xi (t)
-
O\left(\delta\Pba,\delta\Pba^* \right).
\end{eqnarray}
%
Here the first two terms on the r.h.s. of eq.
(\ref{(G-G0-Lo)-var-V}) result from the $\delta \Ga^0$ in the
$\ln$-term of $\Gamma$, these are derived very similarly to that
in eq. (\ref{G-G0-var-mu}) by using eqs. (\ref{dG-free/V}) and
(\ref{dS/V}) for the $\delta \Ga^0$ variation.
The third term comes from the variation of internal integrations
(\ref{V->V'}) in the $\Phi$-functional. Deriving it, we  used the
fact that the variation of internal integrations in the
$\Phi$-functional is equivalent to its variation in $\lambda$ (cf.
eqs. (\ref{V->lambda}) and (\ref{d-lambda/V})). The term
$O\left(\delta\Pba,\delta\Pba^* \right)$ denotes the contribution
of variations of mean fields, which is explicitly cancelled out by
the respective variations in eq. (\ref{Lo-var-V}).

Collecting all the above terms of eqs. (\ref{G0-var-V}),
(\ref{Lo-var-V}) and (\ref{(G-G0-Lo)-var-V})  and using Dyson's
equation (\ref{Dyson}), we arrive at
%
\begin{eqnarray}
\label{G-var-V} \delta \Gamma &=&  \oint \di x 3 \xi (t) \left[
\vphantom{\oint}
\frac{1}{6} \ii
\kappa \left[
\left( \widehat{\vec{p}}_x + (\widehat{\vec{p}}_y)^* \right)^2 \Ga
(x,y)\right]_{y=x}  \right.\nonumber
\\  &-& \left.
\frac{1}{6}
\kappa \left[\left(\nabla_x - \nabla_y \right)^2 \Pba^*(x)
\Pba(y)\right]_{y=x}
-
\left({\cal E}^{\scr{int}}(x) -{\cal E}^{\scr{pot}}(x)\right)
\right]\,.
\end{eqnarray}
%
Comparing the r.h.s. of this expression with that for the
energy--momentum tensor (\ref{E-M-G-tensor}) and taking account of
eq. (\ref{V->V'}), we see that
%
\begin{eqnarray}
\label{G-var-V1} \delta \Gamma =  \oint \di t \frac{1}{3}
\vphantom{\oint}
\Theta^{ii}
\delta V(t),
\end{eqnarray}
%
with the summation over the repeated Latin  indices $i=1,2,3$.
Thus,
%
\begin{eqnarray}
\label{G-var-V11} \delta \widetilde{\Omega} =   -\frac{1}{3}
\vphantom{\oint}
\Theta^{ii}
\delta V(t).
\end{eqnarray}
%
Combining (\ref{O-var-mu}) and (\ref{G-var-V11}) we arrive at the 
expression
\begin{eqnarray}
\label{G-var-V11hom}
\widetilde{\Omega} =   -\frac{1}{3} \int
\di^3x  \vphantom{\oint}
\Theta^{ii}
 .
\end{eqnarray}
Note that  deriving above results we have not assumed anywhere the
thermal equilibrium. Therefore, they hold for arbitrary
non-equilibrium.

In equilibrium, in the rest frame, where collective velocity of
the matter is zero, $\Theta^{ii}/3=P_{\rm eq}$  ($P_{\rm eq}$
is the equilibrium pressure of the matter). Therefore, we can
associate $\widetilde{\Omega}$ with a non-equilibrium pressure
$\widetilde{P}$:
\begin{eqnarray}
\label{P-noneq} \widetilde{\Omega} =   - \int \di^3x
\vphantom{\oint} \widetilde{P},
\end{eqnarray}
of course, in the local rest frame of the matter.  In general,
local rest frames are different at different ${\bf x}$. Therefore,
here and below integration over $\di^3x$ should be understood as
the integration over small volume around ${\bf x}$.

Slightly out of
equilibrium the energy--momentum tensor takes the form \cite{LL06}
%
\begin{eqnarray}
\label{EM-tens-eq}
\Theta^{\mu\nu} &=& \left(\varepsilon_{\rm eq} +P_{\rm eq}\right) u^\mu u^\nu - g^{\mu\nu} P_{\rm eq} + \pi^{\mu\nu}
\\
\pi^{\mu\nu} &=&   \eta
\left(
\partial^{\mu}u^{\nu} + \partial^{\nu}u^{\mu}
-u^{\mu} u_{\lambda}\partial^{\lambda}u^{\nu}
-u^{\nu} u_{\lambda}\partial^{\lambda}u^{\mu}
\right)
+\left(\zeta - \frac{2}{3} \eta \right)
\left(g^{\mu\nu} -u^{\mu} u^{\nu} \right)\partial_{\lambda}u^{\lambda}
\end{eqnarray}
%
where $\varepsilon_{\rm eq}$ and $P_{\rm eq}$ are local
equilibrium proper energy density and   pressure, respectively,
and $u_\mu $ is a local 4-velocity. All these quantities are, in
general, $x$ dependent. Non-equilibrium effects are associated
with $\pi^{\mu\nu}$ tensor, i.e. with shear ($\eta$) and bulk
($\zeta$) viscosities. Here we put aside various causal extensions
of dissipative hydrodynamics \cite{Romatschke:2009im}. Then the
non-equilibrium potential takes the form
%
\begin{eqnarray}
\label{Om-sl-n-eq}
\widetilde{\Omega} =   -\frac{1}{3} \int
\di^3x  
\left[
\left(\varepsilon_{\rm eq} +P_{\rm eq}\right) u^i u^i + 3P_{\rm eq}
+  2\eta
\left(
\partial^{i}u^{i}
-u^{i} u_{\lambda}\partial^{\lambda}u^{i}
\right)
-
\left(\zeta - \frac{2}{3} \eta \right)
\left(3 +u^{i} u^{i} \right)\partial_{\lambda}u^{\lambda}
\right].
\end{eqnarray}
%
In the local rest frame, i.e. at ${\bf u}=0$, it reads
%
\begin{eqnarray}
\label{Om-sl-n-eq-sm-u}
\widetilde{\Omega} =   - \int
\di^3x  
\left[
P_{\rm eq}
-
\zeta
\partial_{i}u^{i}
\right],
\end{eqnarray}
%
which is a key relation for evaluation of $\zeta $.  To obtain the latter
equality we used the fact that the term $\propto \partial_{0}u^{0}$ in eq.
(\ref{Om-sl-n-eq-sm-u}) should be  omitted because
$
\partial_{0} u^{0}\propto u^{i}
\partial_{0} u^{i}$. Since $\zeta >0$ for stable systems, the
non-equilibrium contribution to the pressure
(\ref{Om-sl-n-eq-sm-u}) is negative for expanding system (for
$\partial_i u^i >0$) and positive for contracting system (for
$\partial_i u^i <0$), see \cite{V2011}.

It is reasonable to use the following setup in order to evaluate
the bulk viscosity. Proceeding from certain $\Phi$ functional, a
problem of  Hubble-like expansion of the matter should be solved
within the non-equilibrium Green's function technique. The
advantage of the Hubble-like setup is that the system remains
spatially homogeneous during the expansion. The ``rate'' of 
expansion, $\partial_{i}u^{i}$, is an external parameter, that
can be prescribed any value because the problem does not contain
any spatial scale.  Initial conditions for this expansion
should be chosen in such a way that deviations from the
corresponding equilibrium solution are small. A possible choice is
the equilibrium solution boosted in accordance with the Hubble
rule. This boosted equilibrium initial condition still presents a
{\em locally} equilibrium configuration. Genuine non-equilibrium
develops only during the expansion governed by real-time-contour
equations of motion.  Upon solving this problem, the
non-equilibrium  pressure $\widetilde{P}=P_{\rm eq} -\zeta
\partial_{i}u^{i}$ can be determined in terms of 2PI
diagrams according to rules formulated in subsect. \ref{Diagrams}.
The equilibrium pressure $P_{\rm eq}$ should be determined
within a
thermodynamic calculation based on the same $\Phi$ functional.
Then, the bulk viscosity can be directly determined.

\section{Summary}

 We constructed a non-equilibrium potential, associated with a
pressure for non-equilibrium systems and formulated the rules for
its calculation within the $\Phi$-functional method.
 The non-equilibrium potential transforms to the ordinary thermodynamic
potential in case of the local equilibrium and equilibrium
systems.
By means of variations of this non-equilibrium potential over respective
parameters it is possible to calculate the same quantities in the non-equilibrium
as those with the help of the thermodynamic potential.
Since our non-equilibrium potential can be expressed in
terms of 2PI diagrams within
the non-equilibrium Green's function technique on the Schwinger-Keldysh contour, it opens
new possibilities of evaluating these quantities. In particular,
a possible application of this method to calculation of the bulk
viscosity is described.

\section*{Acknowledgements}
We are grateful to  J. Knoll for numerous discussions and valuable
remarks.  Y.B.I. was partially supported  by the grant
NS-215.2012.2.

\appendix
\section{Contour Function Relations}
\label{discontinouity}
Due to the change of operator ordering genuine multi-point functions are
discontinuous in general, when two contour coordinates become identical. In
particular, two-point functions like $\ii F(x,y)=\left<\Tc {\widehat
    A(x)}\medhat{B}(y)\right>$ become
%
\begin{eqnarray}\label{Fxy}
\hspace*{-1cm}\ii F(x,y) &=&
\left(\begin{array}{ccc}
\ii F^{--}(x,y)&&\ii F^{-+}(x,y)\\[3mm]
\ii F^{+-}(x,y)&&\ii F^{++}(x,y)
\end{array}\right)=
\left(\begin{array}{ccc}
\left<{\cal T}\medhat{A}(x)\medhat{B}(y)\right>&\hspace*{5mm}&
\left<\medhat{B}(y)\medhat{A}(x)\right>\\[5mm]
\left<\medhat{A}(x)\medhat{B}(y)\right>
&&\left<{\cal T}^{-1}\medhat{A}(x)\medhat{B}(y)\right>
\end{array}\right),
\end{eqnarray}
%
where ${\cal T}$ and ${\cal T}^{-1}$ are the usual time and
anti-time ordering operators.   From eq. (\ref{Fxy}) follow
relations between non-equilibrium and usual retarded and advanced
functions
%
\begin{eqnarray}\label{Fretarded}
F^R(x,y)&=&F^{--}(x,y)-F^{-+}(x,y)=F^{+-}(x,y)-F^{++}(x,y)\nonumber\\
&:=&\Theta(x_0-y_0)\left(F^{+-}(x,y)-F^{-+}(x,y)\right),\nonumber\\
F^A (x,y)&=&F^{--}(x,y)-F^{+-}(x,y)=F^{-+}(x,y)-F^{++}(x,y)\nonumber\\
&:=&-\Theta(y_0-x_0)\left(F^{+-}(x,y)-F^{-+}(x,y)\right),
\end{eqnarray}
%
where $\Theta(x_0-y_0)$ is the step function of the time difference.

Discontinuities of a two-point function may cause problems for
derivatives, in particular, since they often occur simultaneously
in products of two or more two-point functions. The proper
procedure is, first, with the help of eq. (\ref{Fretarded}) to
represent the discontinuous parts in $F^{--}$ and $F^{++}$ by the
continuous $F^{-+}$ and $F^{+-}$ times $\Theta$-functions, then to
combine all discontinuities, e.g. with respect to $x_0-y_0$, into
a single term proportional to $\Theta(x_0-y_0)$, and taking
derivatives. One can easily check that in the two particularly
relevant cases
%
\begin{eqnarray}
\label{diffrules0}
\label{diffrules}
&&
\frac{\partial}{\partial x_{\mu}}
\oint\di z\left(
F(x^i,z)G(z,x^j)+G(x^i,z)F(z,x^j)\right),
\\ \label{diffrules1}
&&\left[\left(\frac{\partial}{\partial x_{\mu}}
              -\frac{\partial}{\partial y_{\mu}}\right)
\oint\di z \vphantom{\frac{\partial}{\partial x_{\mu}}
              -\frac{\partial}{\partial y_{\mu}}}
\left(F(x^i,z)G(z,y^j)-G(x^i,z)F(z,y^j)\right)\right]_{x=y}
\end{eqnarray}
%
 all discontinuities exactly cancel. Thereby the values are
independent of the placement of $x^i$ and $x^j$ on the contour,
i.e. the values are only a function of the physical coordinate
$x$.


>From (\ref{Fretarded}), using the Kubo--Martin--Schwinger
condition \cite{Kubo} for two-point functions in energy-momentum
representation
%
\begin{eqnarray}
\label{KMS-G} F^{-+}(p) = F^{+-}(p) e^{-p_0/T}, \,\,\,\,
\end{eqnarray}
%
one derives the equilibrium form of
two-point  functions
%
\begin{eqnarray}
\label{Geq} F(p) = \left(\begin{array}{ccc} F^R(p)- \ii n(p_0)
F^{\rm sp} (p) &\hspace*{5mm}& -\ii n(p_0)F^{\rm sp} (p)
\\[3mm] -\ii \left(1+ n(p_0)\right) F^{\rm sp} (p) &&
-F^A(p)- \ii n(p_0) F^{\rm sp} (p)
\end{array}\right),
\end{eqnarray}
%
where $F^{\rm sp}(p)=-\ii F^A(p)+\ii F^R(p)=-2\Im F^R(p)$ is the
corresponding spectral function,
$n(p_0)$ is the thermal Bose--Einstein occupation number
%
\begin{equation}\label{occup}
n(p_0) =\left[\exp((p_0 -e\mu)/T)- 1\right]^{-1}\;
\end{equation}
%


\end{document}